\documentclass[12pt]{article}

\usepackage{graphicx}

\begin{document}

\title{Drawing the Free Rigid Body Dynamics According to Jacobi}
\author{Eduardo Pi\~na \\ Department of Physics\\ Universidad Aut\'onoma Metropolitana - Iztapalapa, \\
P. O. Box 55 534 \\ M\'exico, D. F., 09340 Mexico \\
e-mail: pge@xanum.uam.mx}

\date{ }

\maketitle

\section*{Abstract}

Guided by the Jacobi's work published the year before his death about the rotation of a rigid body, the behaviour of the rotation matrix describing the dynamics
of the free rigid body is studied. To illustrate this dynamics one draws on a unit sphere the trace of the three unit vectors, in the body system along the
principal directions of inertia. A minimal set of properties of Jacobi's elliptic functions are used, those which allow to compute with the necessary precision
the dynamics of the rigid body without torques, the so called Euler's top. Emphasis is on the paper published by Jacobi in 1850 on the explicit expression for
the components of the rotation matrix. The tool used to compute the trajectories to be drawn are the Jacobi's Fourier series for {\sl theta} and {\sl eta}
functions with extremely fast convergence. The Jacobi's {\sl sn}, {\sl cn} and {\sl dn} functions, which are better known, are used also as ratios of {\sl theta}
functions which permit quick and accurate computation. Finally the main periodic part of the herpolhode curve was computed and graphically represented.
\newpage

\section{Introduction}
The history of the dynamics of the free asymmetric rigid body started with Euler. In the Whittaker's book of mechanics one finds several references to the
original papers \cite{1}.

At present time, the dominant source for a detailed and profound treatment of the dynamics of the rigid body is the Klein and Sommerfeld treatise, recently
translated into English \cite{2}.

The purpose to contributing to this venerable subject is the drawing of the predicted motion of the body represented by the trace of the three unit vectors,
forming the rows of the rotation matrix, portraying on the unit sphere close trajectories representing the main periodic behaviour of the rigid body.  Actually,
the third row coincides with the representation of the angular momentum vector as is seen from the body system in the intersection of the sphere of angular
momentum, and the ellipsoid of constant energy. This trajectory of the angular momentum vector has been graphically represented in different publications, see
for example the Bender and Orzag book \cite{3}.

On the contrary, to my knowledge, the other two rows of the rotation matrix, have not received the pictorial representation of its motion, notwithstanding this
motion for the second row has similar properties of symmetry that the angular momentum whereas the first row has different symmetries that the other two rows. In
any case each of the three trajectories are periodic of the same period. The second and third row are symmetric with respect to the same two coordinate planes in
the body system of principal moments of inertia. The first row, on the contrary, is skew symmetric with respect to the same planes. In my opinion this behaviour
is an essential knowledge of the physics of the free rotation in space far from acting torques, which deserves to be known and taken in account.

In order to present those facts and draw accurately as needed those curves, one uses a minimum of properties of elliptic functions, just those which are
efficient and precise. Everything supported by the magistral Jacobi's work.

Our purpose is to dilute the embarrassment expressed by the historian of mathematics E. T. Bell \cite{4} who writes:

"The rotation of the rigid body, for example, yields numerous elegant exercises in the elliptic theta functions;
 but few engineers who must busy themselves with rotation have time for elegant analysis."

 After Jacobi, the excellent mathematician K. W. T. Weierstrass contributed in extraordinary form to the theory of the elliptic functions. In the cited treatises
 both in Whittaker as in Klein and Sommerfeld, the properties of the new functions introduced by Weierstrass are used. However in this paper almost nothing of
 the Weierstrass functions are included.

The basis to ignore the Weierstrass' work, in this document, is justified by the following text extracted from the Whittaker and Watson treatise of Analysis
\cite{5} which refers to the mathematics of the rigid body dynamics as:

"This result determines the mean precession about the invariable line in the motion of the rigid body relative to its center of gravity under forces whose
resultant passes through its center of gravity. It is evident that, for purposes of computation, a result of this nature is preferable to the corresponding
result in terms of Sigma-functions and Weierstrassian Zeta-functions, for the reason that the Theta-functions have a specially simple behaviour with respect to
the real period--the period which is of importance in Applied Mathematics--and that the $q$-series are much better adapted for computation than the product by
which the Sigma-function is most simply defined."

 The previous text from two recognized experts in elliptic functions supports the fact that for computing and also for drawing accurately our tools are the best.
 One can be more explicit: the {\sl theta} functions used here are written in terms of Fourier series with $n$-coefficients which are proportional to a number
 $q$, lower than 1, with an exponent $n^2$. In such computations less than ten terms are necessary to obtain a precision of 15 places, for numbers $q$ different
 from 1 in 1 over a million, one exceptional case, without general interest. One should use the {\sl theta} functions because its extremely fast convergence.

 Richard Bellman confirms this point without any doubt in his brief book on {\sl theta} functions \cite{6}.

 The main objective of this paper is to represent by drawings the motion of a rigid body formed by particles which relative mutual distances do not change in
 time. The center of mass is assumed fixed at the origin of coordinates in the inertial system. The position of particle $i$ of the body in the inertial system
 ${\bf r}_i$ is known in terms of the entries of the rotation matrix, because the existence of a coordinate system fixed to the body which will be named the body
 system, with the same origin of coordinates as the inertial system, at the rest center of mass, where all the coordinates of the rigid body ${\bf a}_i$ are
 constants of motion. The rotation matrix $\bf R$ transforms the coordinates from the body system to the inertial system
\begin{equation}
{\bf r}_i = {\bf R a}_i \, , \quad \mbox{(for all $i$).}
\end{equation}

In what follows we will be concentrated in the drawing of the components of the rotation matrix, which to maintain the constant distance between particles,
requires its rows to be formed by the components of three unit vectors, mutually orthogonal, that form a right tern, namely, the $\times$ product of the two
first vectors of $R$ is equal to the third row of the same. The ortho-normality of the rows of the rotation matrix is represented in matrix notation as
\begin{equation}
{\bf R}^{\rm T} {\bf R} = {\bf R} {\bf R}^{\rm T} = {\bf E} \, ,
\end{equation}
where ${\bf E}$ is the unit matrix. The super-index T to the right of a vector or matrix denotes the transposed vector or matrix. Therefore one has the inverse
matrix of the rotation matrix is equal to its transposed matrix.

 The derivative with respect to time is denoted with a point on the function to be derived. The derivative with respect to time of the equations in (2), uncovers
 the existence of the antisymmetric matrices defining the angular velocity
$$
\mbox{\boldmath $\omega$} \times = \left( \begin{array}{ccc}
0 & -\omega_3 & \omega_2 \\
\omega_3 & 0 & -\omega_1 \\
-\omega_2 & \omega_1 & 0
\end{array} \right) \equiv {\bf R}^{\rm T} \dot{\bf R} \, ,
$$
\begin{equation}
\mbox{\boldmath $\Omega$} \times = \left( \begin{array}{ccc}
0 & -\Omega_3 & \Omega_2 \\
\Omega_3 & 0 & -\Omega_1 \\
-\Omega_2 & \Omega_1 & 0
\end{array} \right) \equiv \dot{\bf R} {\bf R}^{\rm T} \, .
\end{equation}

The first definition corresponds to the angular velocity in the body system
\begin{equation}
\mbox{\boldmath $\omega$} = \left( \begin{array}{c}
\omega_1 \\
\omega_2 \\
\omega_3
\end{array} \right)\, ,
\end{equation}
which gives the derivative with respect to time of the rotation matrix in the useful form
\begin{equation}
\dot{\bf R} = {\bf R} \, \mbox{\boldmath $\omega$} \times \, .
\end{equation}

The notation with the $\times$ product is used because the product by the right of the antisymmetric matrix with any vector is equal to the  $\times$ product of
the vector of the matrix with the same vector
\begin{equation}
\mbox{\boldmath $\omega$} \times \left( \begin{array}{c}
a_1 \\
a_2 \\
a_3
\end{array} \right) = \left( \begin{array}{ccc}
0 & -\omega_3 & \omega_2 \\
\omega_3 & 0 & -\omega_1 \\
-\omega_2 & \omega_1 & 0
\end{array} \right) \left( \begin{array}{c}
a_1 \\
a_2 \\
a_3
\end{array} \right) \, .
\end{equation}

The angular velocity in the inertial system {\boldmath $\Omega$} will be important in the last part of this work because one pretends to draw the curve, called
herpolhode, portrayed by this vector when one projects it on the orthogonal plane to the angular momentum vector.

From its definitions (3) the angular velocities vectors are related by equation
\begin{equation}
\mbox{\boldmath $\Omega$} \times = {\bf R} \mbox{\boldmath $\omega$} \times {\bf R}^{\rm T} = ({\bf R} \mbox{\boldmath $\omega$} ) \times \, , \quad
\mbox{\boldmath $\Omega$} = {\bf R} \mbox{\boldmath $\omega$}\, .
\end{equation}
where we have used the geometric theorem: the $\times$ product of rotated vectors is equal to the rotation of the $\times$ product of the vectors.

To know the rotation matrix one starts using the conservation of the angular momentum vector. The angular momentum vector (with respect to the center of mass) is
defined by the sum vector of the mass by the $\times$ product of the position and the velocity
$$
{\bf J} = \sum_i m_i {\bf r}_i \times \dot{\bf r}_i = \sum_i m_i ({\bf R a}_i) \times (\dot{\bf R} {\bf a
}_i)
$$
\begin{equation}
= \sum_i m_i ({\bf R a}_i) \times ({\bf R} \mbox{\boldmath $\omega$} \times {\bf a
}_i) = {\bf R} \sum_i m_i {\bf a}_i \times (\mbox{\boldmath $\omega$} \times {\bf a}_i) \, ,
\end{equation}
where the positions were used in terms of the rotation matrix; the time derivative of the rotation matrix as a function of the angular velocity, and the
geometric theorem: the $\times$ product of the rotated vectors is the rotation of the $\times$ product of the vectors. Next the double $\times$ product is used
to obtain
\begin{equation}
{\bf J} = {\bf R}  \sum_i m_i \left[ {\bf a}_i^{\rm T}{\bf a}_i \mbox{\boldmath $\omega$} - {\bf a}_i {\bf a}_i^{\rm T} \mbox{\boldmath $\omega$} \right]= {\bf
R} \sum_i m_i \left[ {\bf a}_i^{\rm T} {\bf a}_i {\bf E} - {\bf a}_i {\bf a}_i^{\rm T} \right] \mbox{\boldmath $\omega$} \, ,
\end{equation}
where one has introduced the unit matrix $\bf E$ to take out as a common factor the angular velocity vector. In this way one finds the angular momentum vector of
the inertial system as the product of the rotation matrix, the inertia matrix of inertia in the body system, and the angular velocity of the same system
\begin{equation}
{\bf J} = {\bf R} \mbox{\boldmath ${\cal I} \omega$}\, .
\end{equation}
The inertia matrix {\boldmath{$\cal I$}} is the following constant symmetric matrix, which is assumed diagonal since the body system could be selected in such a
way.
\begin{equation}
\mbox{\boldmath $\cal I$} = \sum_i m_i \left[ {\bf a}_i^{\rm T} {\bf a}_i {\bf E} - {\bf a}_i {\bf a}_i^{\rm T} \right] = \left( \begin{array}{ccc}
I_1 & 0 & 0 \\
0 & I_2 & 0 \\
0 & 0 & I_3
\end {array} \right) \, .
\end{equation}
 Quantities $I_j$ are positive and are called principal moments of inercia. The body system where the inertia matrix is diagonal is called the system of the
 principal moments of inertia. In what follows all the vectors of the body system have components in this system of coordinates.

\section{Computing the angular momentum in the body system}
 The angular momentum vector is conserved. This is a general theorem of mechanics: if no external torques are present, the angular momentum vector is conserved.
 The inertial system of coordinates is selected so that the angular momentum vector is directed along the coordinate axis 3. The constant magnitude of this
 vector is $J$
\begin{equation}
{\bf J}^{\rm T} = J (0, 0, 1) \, .
\end{equation}

According to equation (10) the components of the angular momentum in the body system are ${\bf L} = ${\boldmath ${\cal I} \omega$}. The magnitude of this vector
is the constant $J$ as it does not change with the rotation, but its direction changes with time. One denotes by $\bf u$ the unit vector in the body system in
the direction of this vector
\begin{equation}
{\bf L} = \mbox{\boldmath ${\cal I} \omega$} = J {\bf u}\, .
\end{equation}

Hence the vector $\bf L$ is rotated into the vector $\bf J$ and the vector $\bf u$ is rotated into the vector $\bf k$:
\begin{equation}
{\bf J} = {\bf R L} \, , \quad {\bf R u} = {\bf k} = \left( \begin{array}{c}
0 \\
0 \\
1
\end{array} \right) \, .
\end{equation}
As the rows of the rotation matrix are mutually orthogonal the second equation in (14) implies that ${\bf u}^{\rm T}$ is the vector forming the third row of the
rotation matrix. One proceeds to compute this vector.

The equation of motion for $\bf L$ is obtained from the derivative with respect to time of the previous equation
\begin{equation}
{\bf 0} = {\bf R} ( \dot{\bf L} + \mbox{\boldmath $\omega$} \times {\bf L}) \, , \quad \dot{\bf u} = {\bf u} \times \mbox{\boldmath $\omega$} \, ,
\end{equation}
where it has been written the time derivative of the matrix $\bf R$ in terms of the angular velocity. This is the Euler equation for the free rigid body.

In other hand, according to (13), the angular velocity is written in terms of vector $\bf L$ as
\begin{equation}
\mbox{\boldmath $\omega$} = \mbox{\boldmath ${\cal I}$}^{-1} {\bf L} \, .
\end{equation}
We find the equation of motion for the components of the angular momentum vector in the body system
\begin{equation}
\dot{\bf L} = {\bf L} \times \mbox{\boldmath ${\cal I}$}^{-1} {\bf L} \, .
\end{equation}
Which has the constants of integration energy $E$ and magnitude of the angular momentum vector
\begin{equation}
{\bf L}^{\rm T} \mbox{\boldmath ${\cal I}$}^{-1} {\bf L} = 2 E \, , \quad {\bf L}^{\rm T} {\bf L} = J^2 \, .
\end{equation}

We turn the attention to the unit vector $\bf u$ in the direction of the angular momentum, dividing the angular momentum vector by its magnitude.

One writes the inverses of the principal moments of inertia and the constant $2 E/J^2$ in the form
\begin{equation}
\frac{1}{I_j} = H + P e_j \, , \quad \frac{2 E}{J^2} = H + P e_0 \, ,
\end{equation}
where $H$ and $P$ are defined in order to lower the number of independent parameters
\begin{equation}
H = \frac{1}{3} \left( \frac{1}{I_1} + \frac{1}{I_2} + \frac{1}{I_3} \right) \, , \quad e_1 + e_2 + e_3 = 0 \,.
\end{equation}
\begin{equation}
P^2 = \frac{4}{9} \left( \frac{1}{I_1^2} + \frac{1}{I_2^2} + \frac{1}{I_3^2} - \frac{1}{I_1 I_2} - \frac{1}{I_2 I_3} - \frac{1}{I_3 I_1} \right) \, , \quad e_1^2
+ e_2^2 +e_3^2 = 3/2 \, ,
\end{equation}
which implies the parameters $e_j$ will be known in terms of one angle $\gamma$
\begin{equation}
e_1 = \cos \gamma \, , \quad e_2 = \cos (\gamma - 2 \pi/3) \, , \quad e_3 = \cos (\gamma + 2 \pi/3) \, .
\end{equation}

The constants of motion that for the angular momentum vector were functions of five parameters $E$, $J$, $I_1$, $I_2$, $I_3$; for the components $u_j$ of the
unit vector in the direction of the angular momentum are written just in terms of the two parameters $\gamma$ and $e_0$
\begin{equation}
u_1^2 + u_2^2 + u_3^2 = 1 \, , \quad e_1 u_1^2 + e_2 u_2^2 + e_3 u_3^2 = e_0 \, .
\end{equation}
These equations suggest to utilize the sphero-conal coordinates $\alpha_1$ and $\alpha_2$ for the components of this vector
$$
u_1 = {\rm cn}(\alpha_2, k_2) {\rm cn}(\alpha_1, k_1) \, ,
$$
$$
u_2 = {\rm dn}(\alpha_2, k_2) {\rm sn}(\alpha_1, k_1) \, ,
$$
\begin{equation}
u_3 = {\rm sn}(\alpha_2, k_2) {\rm dn}(\alpha_1, k_1) \, .
\end{equation}
The sphero-conal coordinates are used at present time. According to the NIST Handbook of Mathematical Functions \cite{7}, the parameter $\beta$ in 29.18.2 of
this handbook corresponds in our notation to $K(k_1) + i K(k_2) - i \alpha_2$, in terms of the function and the arguments which will be defined in this section.
One prefers the form (24) since it is not necessary this far of paper to use complex variable. In the quantum case of the same Euler free rigid body the
Schr\"odinger equation is separated in sphero-conal coordinates \cite{8}. (See for example the R. M\'endez-Fragoso and E. Ley-Koo \cite{9} review paper on
quantum rotations.) The sphero-conal coordinates in the form (24) are also found in the Morse and Feshbach book \cite{10}, as conical coordinates. The
stereoscopic view in 3 dimensions of some of the conical coordinate curves have been drawn in this reference. In our forward figure 1 some coordinate curves
($\alpha_2$ = constant) are drawn on the sphere as the trajectories followed by the angular momentum vector in the system of the principal moments of inertia.

The first equation of (23) is satisfied identically with these coordinates if we use the properties of the Jacobi elliptic functions
\begin{equation}
{\rm sn}^2(\beta, k) + {\rm cn}^2(\beta, k) = 1 \, , \quad k^2 {\rm sn}^2(\beta, k) + {\rm dn}^2(\beta, k) = 1
\end{equation}
and the relation of the constants $k_i$
\begin{equation}
k_1^2 + k_2^2 = 1 \, .
\end{equation}

The second equation of (23) is satisfied identically with these coordinates if $\alpha_2$, $k_1$, $k_2$ are constants chosen with the restrictions
$$
{\rm sn}(\alpha_2, k_2) = \sqrt{\frac{e_1-e_0}{e_1-e_3}} \, , \quad {\rm cn}(\alpha_2, k_2) = \sqrt{\frac{e_0-e_3}{e_1-e_3}} \, , \quad {\rm dn}(\alpha_2, k_2) =
\sqrt{\frac{e_0-e_3}{e_2-e_3}}
$$
\begin{equation}
k_1 = \sqrt{\frac{(e_1-e_2)(e_0-e_3)}{(e_2-e_3)(e_1-e_0)}} \, , \quad k_2 = \sqrt{\frac{(e_1-e_3)(e_2-e_0)}{(e_2-e_3)(e_1-e_0)}} \, , \quad
\end{equation}
 which are redundant because the equations (25) and (26) are identically satisfied. Note that these constants are only functions of the parameters $\gamma$ of
 asymmetry and $e_0$ of energy. For the components $u_1$, $u_2$ and $u_3$ of vector $\bf u$, and of the third row of the rotation matrix $\bf R$, it is
 preferable replace instead of (24)
$$
u_1 = \sqrt{\frac{e_0-e_3}{e_1-e_3}} {\rm cn}(\alpha_1, k_1) \, ,
$$
$$
u_2 = \sqrt{\frac{e_0-e_3}{e_2-e_3}} {\rm sn}(\alpha_1, k_1) \, ,
$$
\begin{equation}
u_3 = \sqrt{\frac{e_1-e_0}{e_1-e_3}} {\rm dn}(\alpha_1, k_1) \, .
\end{equation}

Coordinate $\alpha_1$ is a function of time. To know its behaviour we must know the derivatives of these Jacobi functions. Owed to the quadratic relations (25),
it is sufficient to know one of them, but one includes the three to remark its simplicity and similarity
$$
\frac{d \, {\rm sn}(\beta, k)}{d \, \beta} = {\rm cn}(\beta, k) {\rm dn}(\beta, k) \, ,
$$
$$
\frac{d \, {\rm cn}(\beta, k)}{d \, \beta} = -{\rm sn}(\beta, k) {\rm dn}(\beta, k) \, ,
$$
\begin{equation}
\frac{d \, {\rm dn}(\beta, k)}{d \, \beta} = -k^2 {\rm sn}(\beta, k) {\rm cn}(\beta, k) \, .
\end{equation}

We might know now the equation of motion for the coordinate $\alpha_1$. Substituting (13) in the Euler equation of motion, with the explicit form (28) for the
components of $\bf u$ and the principal moments of inertia in the form (19). One finds the constant derivative
\begin{equation}
\dot{\alpha_1} = P J \sqrt{(e_1-e_0)(e_2-e_3)} \, .
\end{equation}

Note that besides the used parameters $\gamma$ and $e_0$, the time appears without dimensions in the combination $J P t$.

To draw the curve $\bf u$, going over the unit sphere, we should learn to compute the Jacobi elliptic functions. As functions of the real variable $\alpha_1$
they are periodic. The period of the vector $\bf u$ is 4 times the function $K(k_1)$. The function $K(k)$ is a complete elliptic integral of first kind. To
compute it we have the algorithm of the arithmetic-geometric mean \cite{12} allowing to calculate it with rapidity and is described as follows: Take the initial
values $x_0 =1 + k$, $y_0=1 - k$. Iterate $x_n, y_n \rightarrow x_{n+1}=(x_n+y_n)/2, y_{n+1} = \sqrt{x_n y_n}$, up to the desired precision the two averages are
equal to a certain value $x$. Then $K(k) = \pi/ (2 x)$. To draw the vector $\bf u$ one divides the period in equal parts and one draws the components of $\bf u$
calculated in all the points corresponding to the increments of the division of the period.

\begin{figure}

\hfil \scalebox{0.5}{\includegraphics{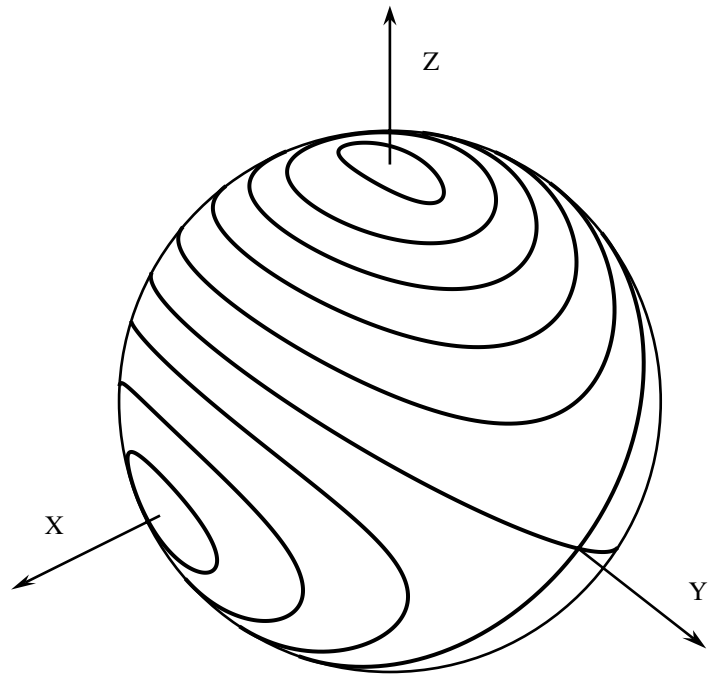}} \hfil

\caption{The sphere of angular momentum intersected by ellipsoids of constant energy. The curves follow the motion of the angular momentum vector on the sphere
of angular momentum for different values of the energy. These curves are symmetric with respect to two of the coordinate planes.}

\end{figure}

We have several efficient algorithms to compute numerically the Jacobi functions \cite{12}. In this paper one uses the Jacobi formula \cite{11} as the ratio of
$\Theta(\beta, k)$ and $H(\beta, k)$ functions, which has the advantage of a very fast convergence, comparable in precision and fastness with other methods, and
the bonus of being one of them, indispensable to the calculus of the other six entries of the rotation matrix $\bf R$.

The {\sl theta} Jacobi functions require other function of the parameter $k$ defined in terms of $K(k)$ by
\begin{equation}
q(k) = \exp(-\pi K(\sqrt{1-k^2})/K(k) \, ,
\end{equation}
which is a positive number lower than one, which is shortened as $q$. The {\sl theta} functions are computed then \cite{11} as Fourier series of real period $2
K(k)$
\begin{equation}
\Theta(\beta,k) = 1 + 2 \sum_{n=1}^\infty (-1)^n q^{n^2} \cos\frac{\pi n \beta}{K(k)} \, ,
\end{equation}
and period $4 K(k)$
\begin{equation}
H(\beta,k) = 2 \sum_{n=0}^\infty (-1)^n q^{(n+1/2)^2} {\rm sen}\, \frac{\pi (2 n+1) \beta}{2 K(k)} \, .
\end{equation}
 These series should be computed up to the necessary precision when the next term is negligible, because the fast convergency to zero of the factor $q^{n^2}$.
 These {\sl theta} functions are the algorithms to compute other elliptic Jacobi functions by means of the equations \cite{11}
\begin{equation}
{\rm sn}(\beta,k) = \frac{H(\beta,k)}{\sqrt{k} \; \Theta(\beta,k)} \, , \quad {\rm cn}(\beta,k) = \frac{\sqrt[4]{1-k^2}}{\sqrt{k}}
\frac{H(\beta+K(k),k)}{\Theta(\beta,k)}
\end{equation}
\begin{equation}
{\rm dn}(\beta,k) = \sqrt[4]{1-k^2} \, \frac{\Theta(\beta+K(k),k)}{\Theta(\beta,k)} \, .
\end{equation}

Jacobi \cite{11} provides the entries of the rotation matrix as ratios of these {\sl theta} functions. He understood the computational advantage of their
functions. In this work one has adopted some priority to the Jacobi's elliptic functions sn, cn, dn whenever it is possible because they are better known
functions and their properties. But the Jacobi's {\sl theta} functions are actually used as algorithms to compute the previous functions. Comparing the Jacobi's
expressions \cite{11} for the third row of the rotation matrix we can observe the same components with the same functions. We are following Jacobi very closely.

As an example of the use of this algorithm to draw on the angular momentum sphere the intersections with ellipsoids of constant energy one has Fig. 1,  in which
the asymmetry parameter was selected as $\gamma = \pi/5$ and several energy parameters of energy $e_0$. This is the trajectory followed by the unit vector along
the third row of the rotation matrix on the sphere of unit radius. The observer of the picture has the spherical coordinates $2 \pi/7, 2 \pi/7$ with respect to
the principal inertia directions.

\section{The entries perpendicular to the body angular momentum in the rotation matrix}
In this section one analyses the paleography of the rotation matrix as a function of time, of the motion without torques of a rigid body, as published by C. G.
J. Jacobi \cite{11} at 1850. Understood paleography as the study of an old text and its traduction or explanation in modern terms. The nine components of the
rotation matrix were written by Jacobi by means of a set of {\sl theta} functions, the same functions used in the previous section. The third row of the rotation
matrix, one pointed before, was written in sphero-conal coordinates in terms of the $H$ and $\Theta$
{\sl theta} functions. Jacobi finds the other six components of the same matrix expressed also by using ratios of the same functions, although now with complex
arguments. The corresponding entries in the two first rows of the rotation matrix involves three {\sl theta} functions, repeated with different argument, and
multiplied, by the real and the other by the imaginary part of a {\sl theta} function with argument the complex variable $\alpha_1 + i \alpha_2$. Remember
$\alpha_1$ is proportional to time, and $\alpha_2$ is a constant coordinate  satisfying three equations in (27), but its numerical value should be determined
now.

\begin{figure}

\hfil \scalebox{0.3}{\includegraphics{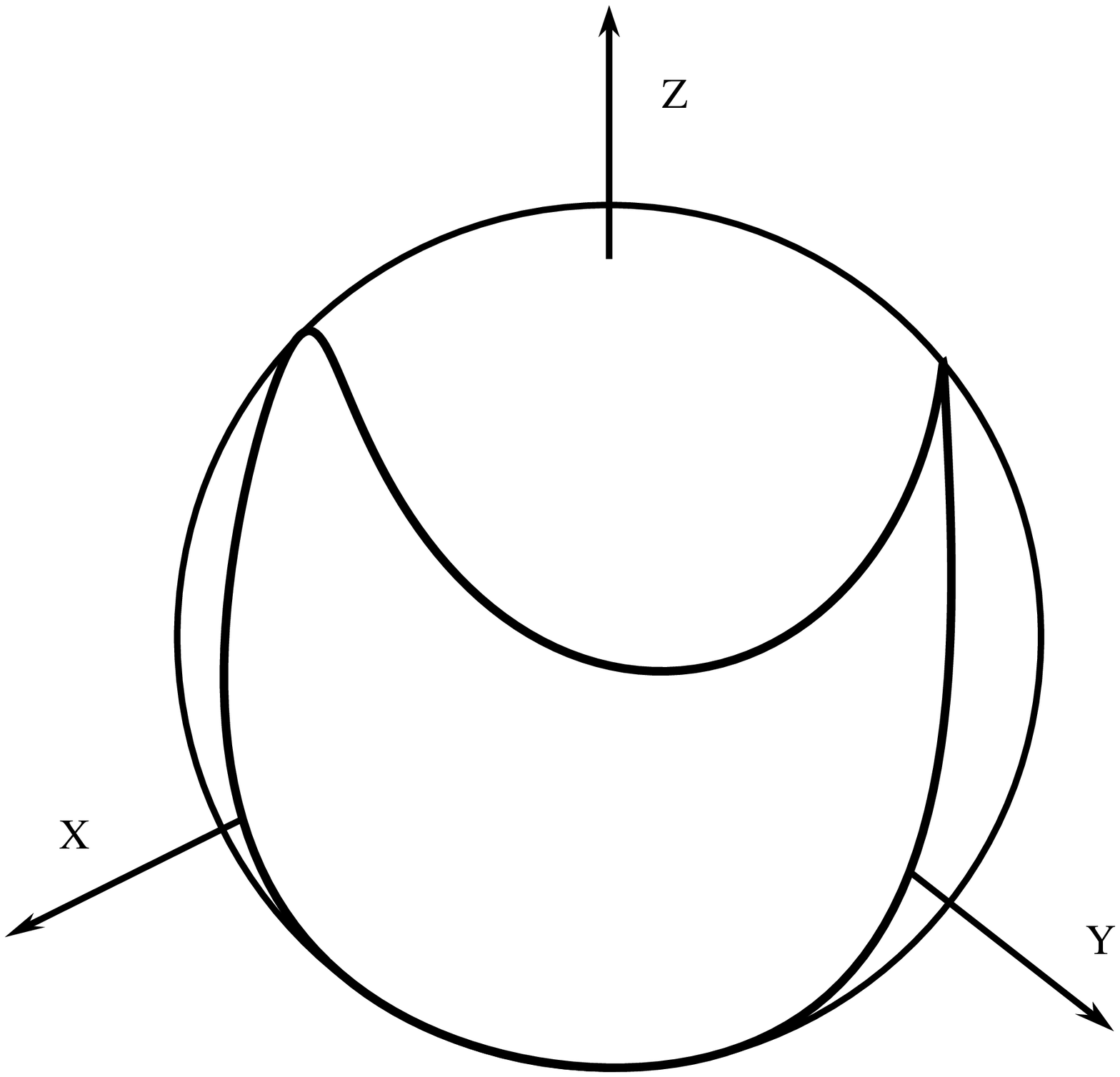}} \hfil \scalebox{0.3}{\includegraphics{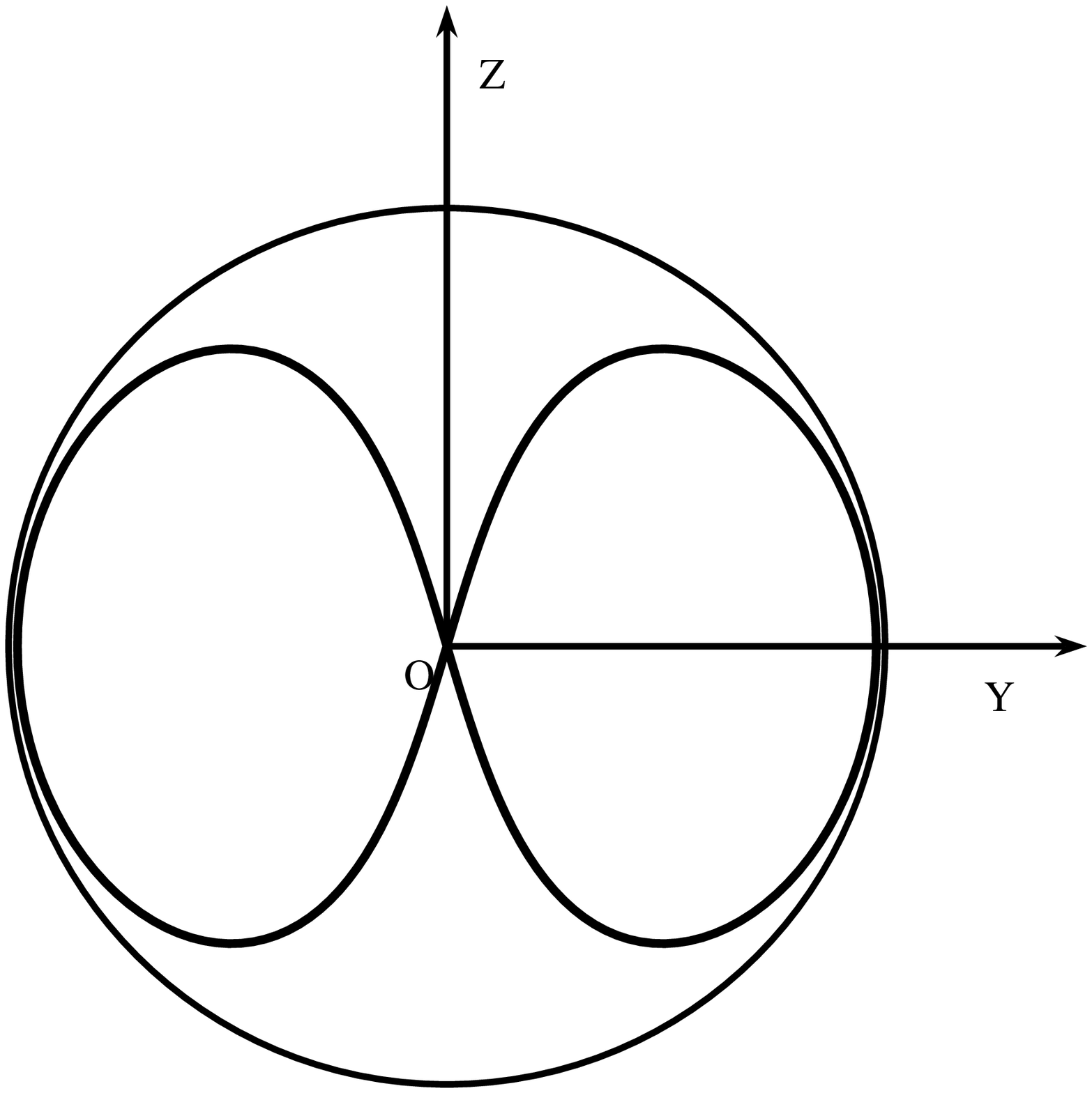}} \hfil
\hfil \scalebox{0.3}{\includegraphics{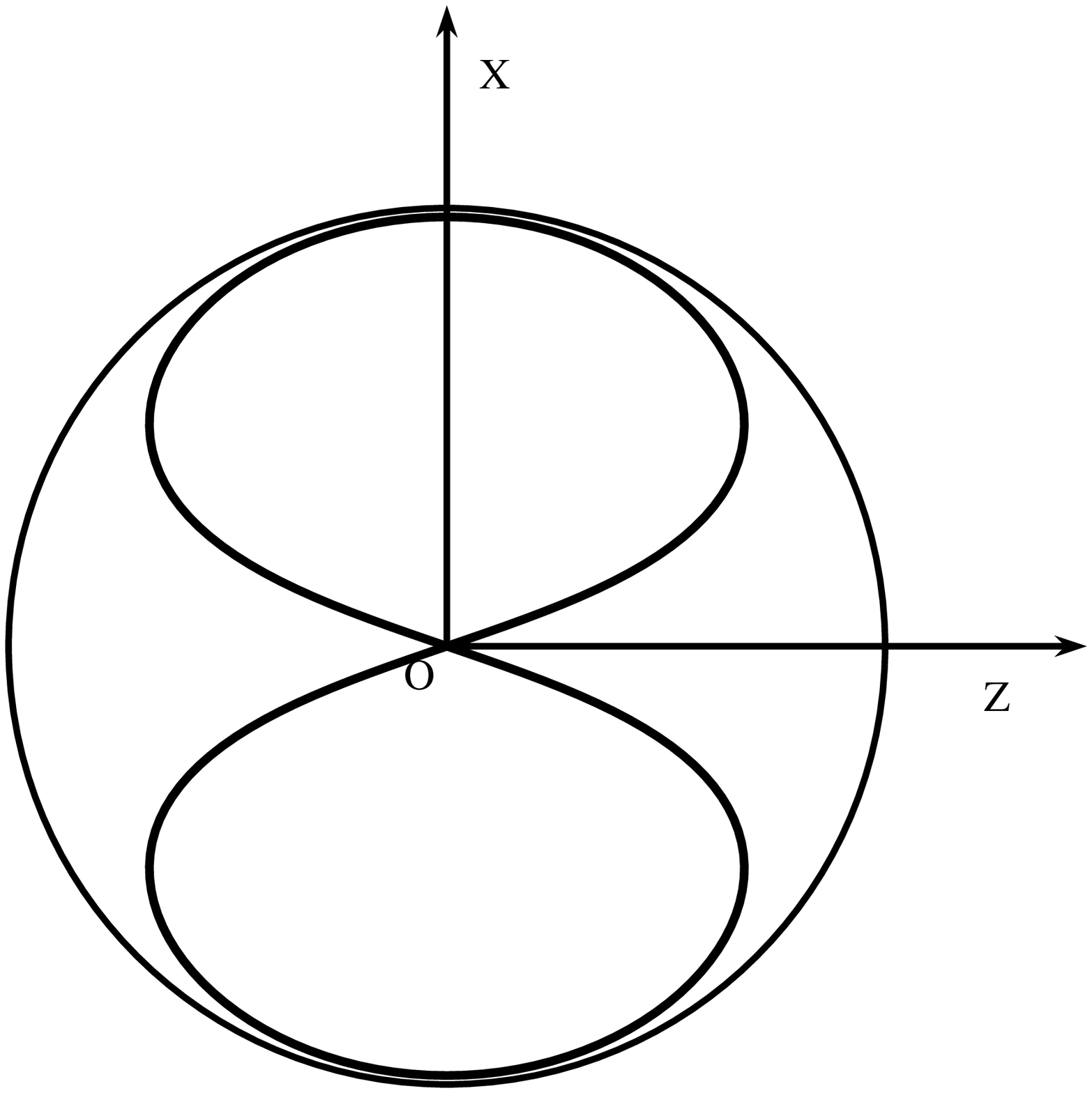}} \hfil \scalebox{0.3}{\includegraphics{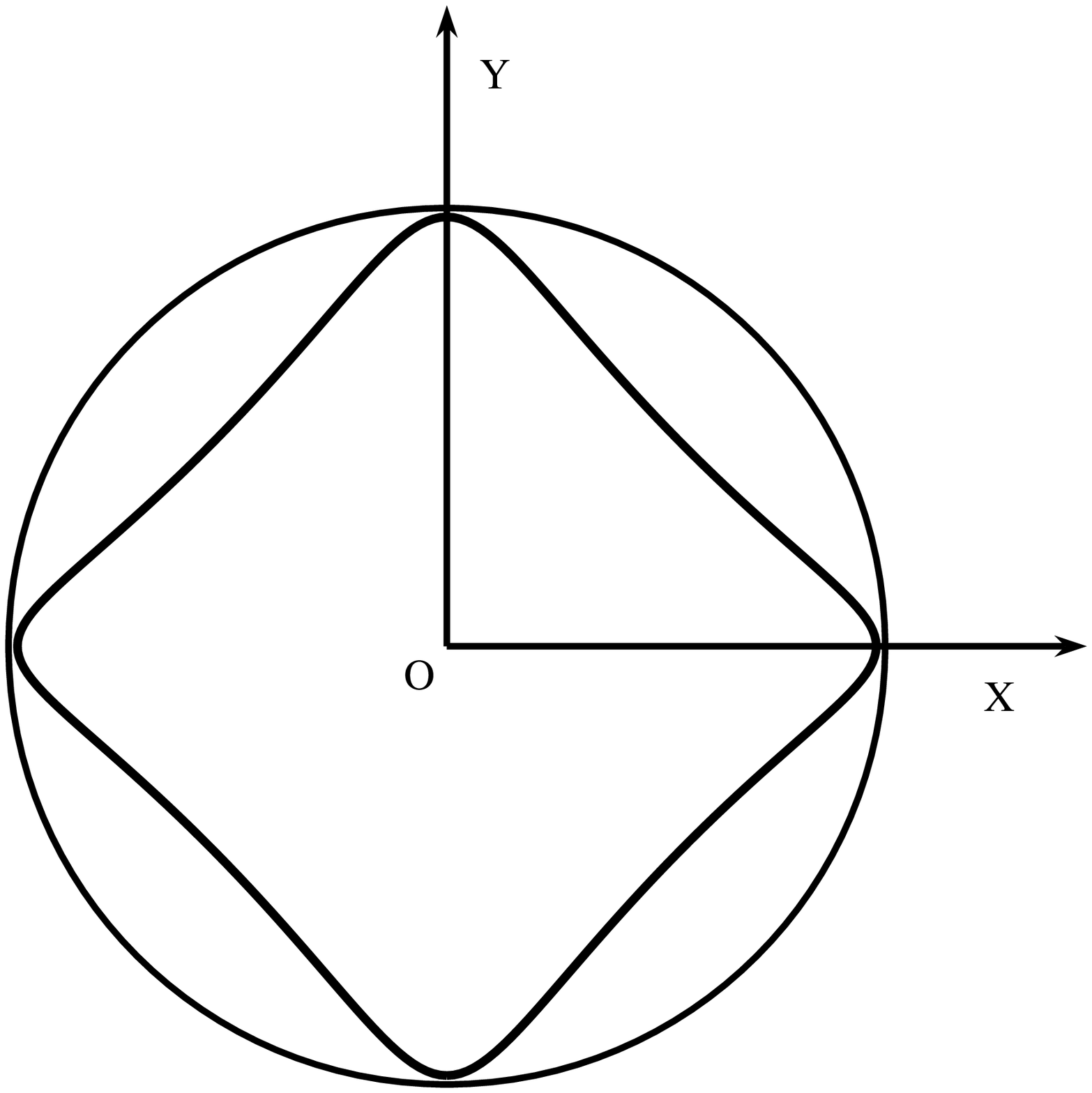}} \hfil

\caption{The trajectory followed by the vector forming the first row of the rotation matrix in a period. Perspective view and projection on the coordinate planes
in the system of principal moments of inertia. The curve is antisymmetric with respect to the two coordinate planes crossing at the OZ axis of coordinates. }

\end{figure}33

Since $\alpha_2$ is an elliptic integral of first kind
\begin{equation}
\alpha_2 = \int_0^{\sqrt{\frac{e_1-e_0}{e_1-e_3}}} \frac{d x}{\sqrt{(1-x^2)(1-k_2^2 x^2)}} \, .
\end{equation}
To attain a more precise value for $\alpha_2$, one introduces the numerical value of this integral as approximated root and look for the root $\alpha_2$ of the
equation
\begin{equation}
{\rm sn}(\alpha_2, k_2) - \sqrt{\frac{e_1-e_0}{e_1-e_3}} = 0 \, .
\end{equation}
Actually one finds two roots for $\alpha_2$ in the interval $[0, 2 K(k_2)]$. The smaller of the two, to be selected, correspond to the choice of positive values
in (27).

The $\Theta$ functions of complex argument are defined by the same Fourier series as before in (32) and (33). The modification due to the imaginary argument
changes the coefficient of the Fourier series, however the convergence remains very fast.

Observing the two first rows of the rotation Jacobi's matrix one deduces the convenience of combining the vector $\bf s$ of the first row as the real part, and
vector $\bf t$ of the second row as minus the imaginary part of a complex vector, function of a complex variable. The two vectors are orthogonal to the third
row, so the complex vector is too. The scalar product of the complex vector with itself should be zero as a consequence of the orthogonality of the real and
imaginary parts and the same magnitude. We call null the vectors with square zero and note is not required the unit magnitude of the real and imaginary vectors
to be null, it is sufficient orthogonality and same magnitude.

It appears a common factor of the complex vector which is written in our notation as
\begin{equation}
{\bf s} - i {\bf t}= \frac{1}{H(i \alpha_2 + K(k_1)) \Theta(\alpha_1)} \left( \begin{array}{c}
\Theta(K(k_1)) H(\alpha_1 + i \alpha_2) \\
 - \Theta(0) H(\alpha_1 + i \alpha_2+K(k_1)) \\
 -i H(K(k_1)) \Theta(\alpha_1 + i \alpha_2)
  \end{array}  \right) \, ,
\end{equation}
where one has suppressed the second argument of the {\sl theta} functions because it is always the same: $k_1$. One has changed also the sign of the first
component because one assumes a Jacobi's mistake to be verified in the sequel.

One notes that components 1 and 2 divided by the third allows to introduce the functions $sn$ and $cn$ of complex argument as the ratio of $theta$ functions,
$$
\frac{\Theta(K(k_1))\, H(\alpha_1 + i \alpha_2)}{H(K(k_1)) \, \Theta(\alpha_1 + i \alpha_2)} = \frac{{\rm sn}\, (\alpha_1 + i \alpha_2, k_1)}{{\rm sn}\, (K(k_1),
k_1)} = {\rm sn}\, (\alpha_1 + i \alpha_2, k_1)\, ,
$$
$$
\frac{\Theta(0)\, H(\alpha_1 + i \alpha_2) + K(k_1))}{H(K(k_1)) \, \Theta(\alpha_1 + i \alpha_2)} = \frac{{\rm cn}\, (\alpha_1 + i \alpha_2, k_1)}{{\rm cn}\, (0,
k_1)} = {\rm cn}\, (\alpha_1 + i \alpha_2, k_1)\, ,
$$
after this, appears a common factor given by the third component, then the complex vector can be written as
\begin{equation}
{\bf s} - i {\bf t} = \frac{H(K(k_1)) \Theta(\alpha_1 + i \alpha_2)}{H(K(k_1) + i \alpha_2) \Theta(\alpha_1)} \left( \begin{array}{c}
{\rm sn}(\alpha_1 + i \alpha_2, k_1) \\
-{\rm cn}(\alpha_1 + i \alpha_2, k_1) \\
-i
\end{array} \right) \, .
\end{equation}
One notes the vector on the right is a null vector as a consequence of the first quadratic identity in (25), although the magnitudes of the real and imaginary
parts of this vector are not equal to 1.

\begin{figure}

\hfil \scalebox{0.3}{\includegraphics{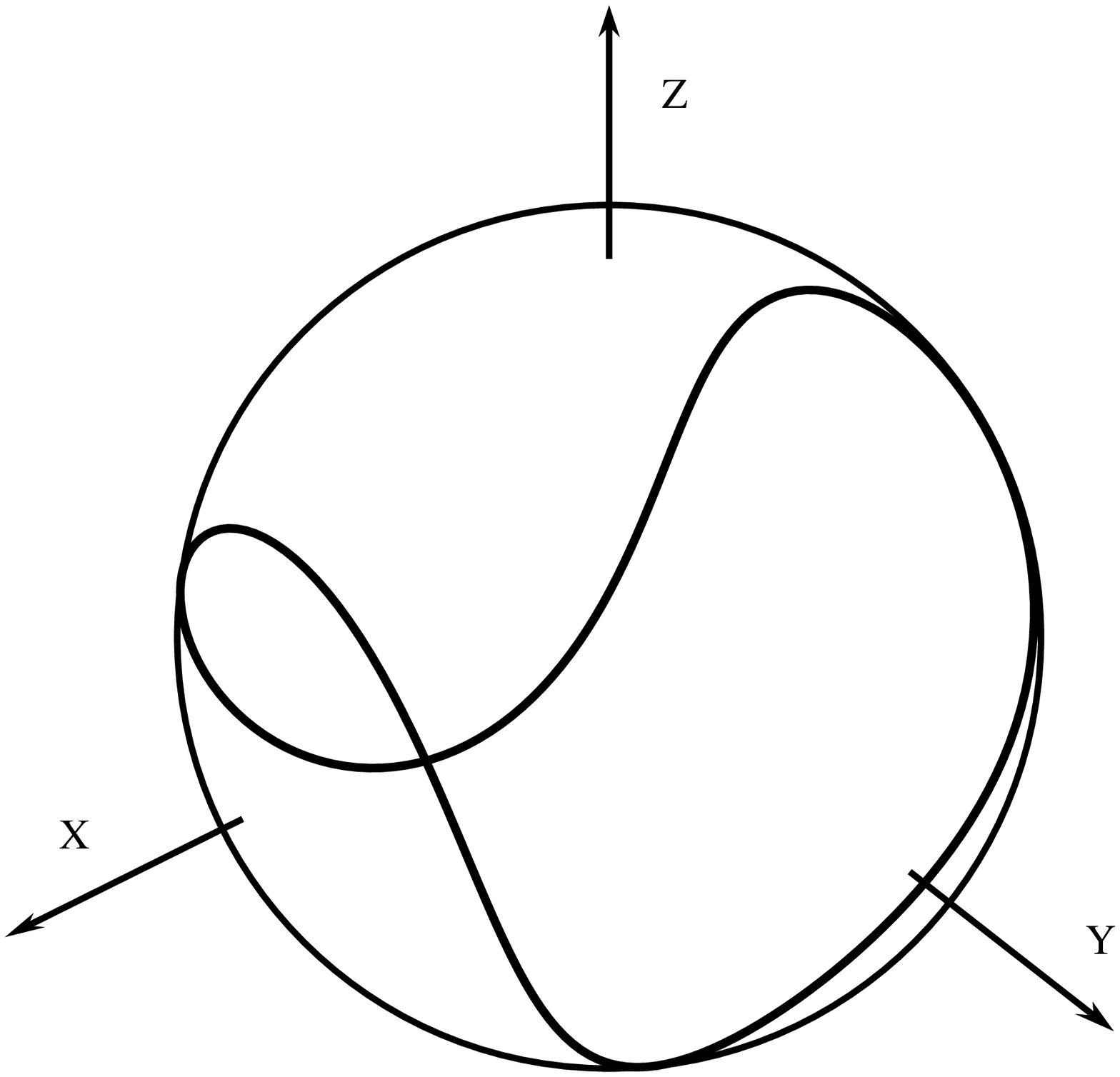}} \hfil \scalebox{0.3}{\includegraphics{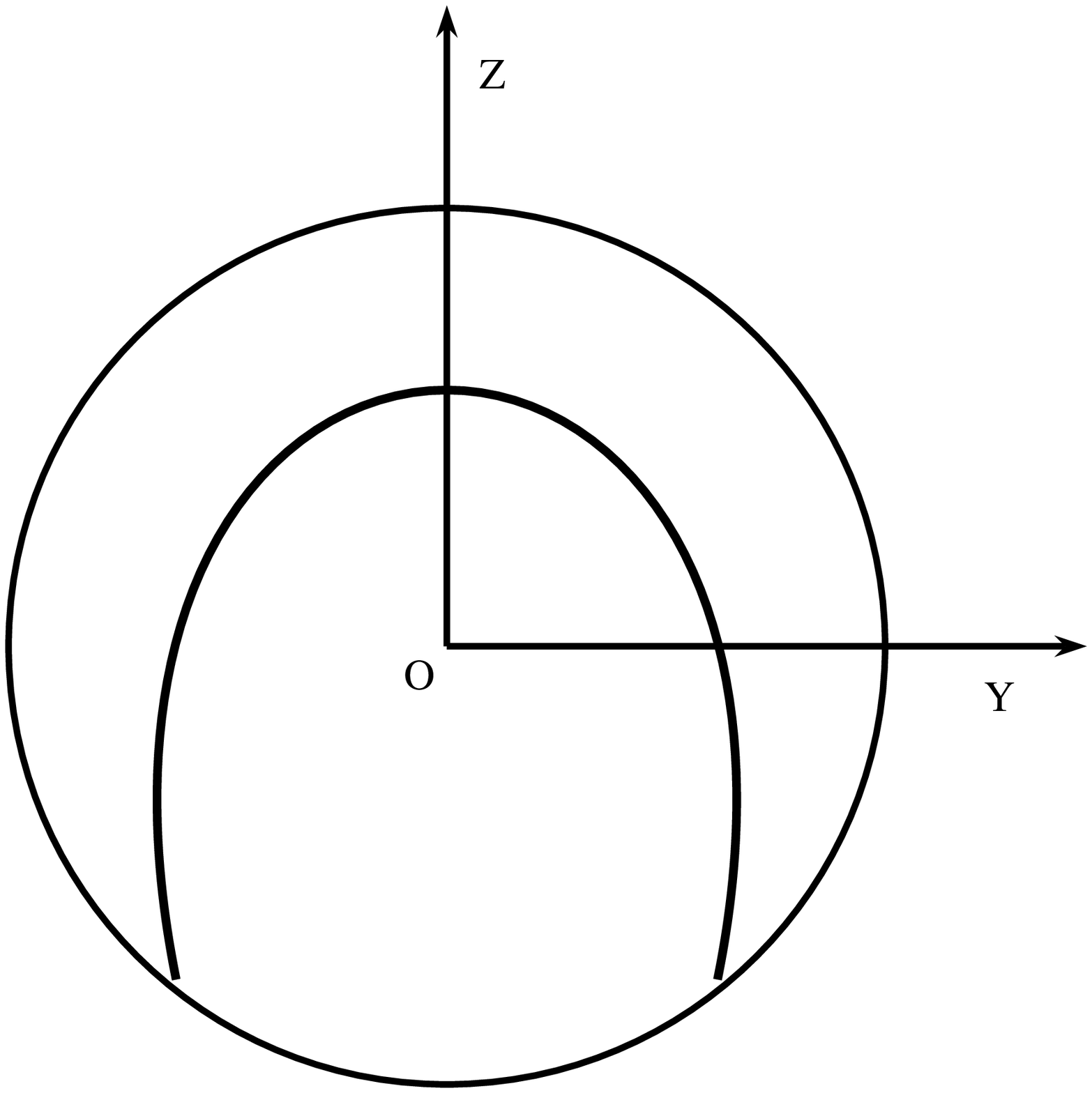}} \hfil

\

\hfil \scalebox{0.3}{\includegraphics{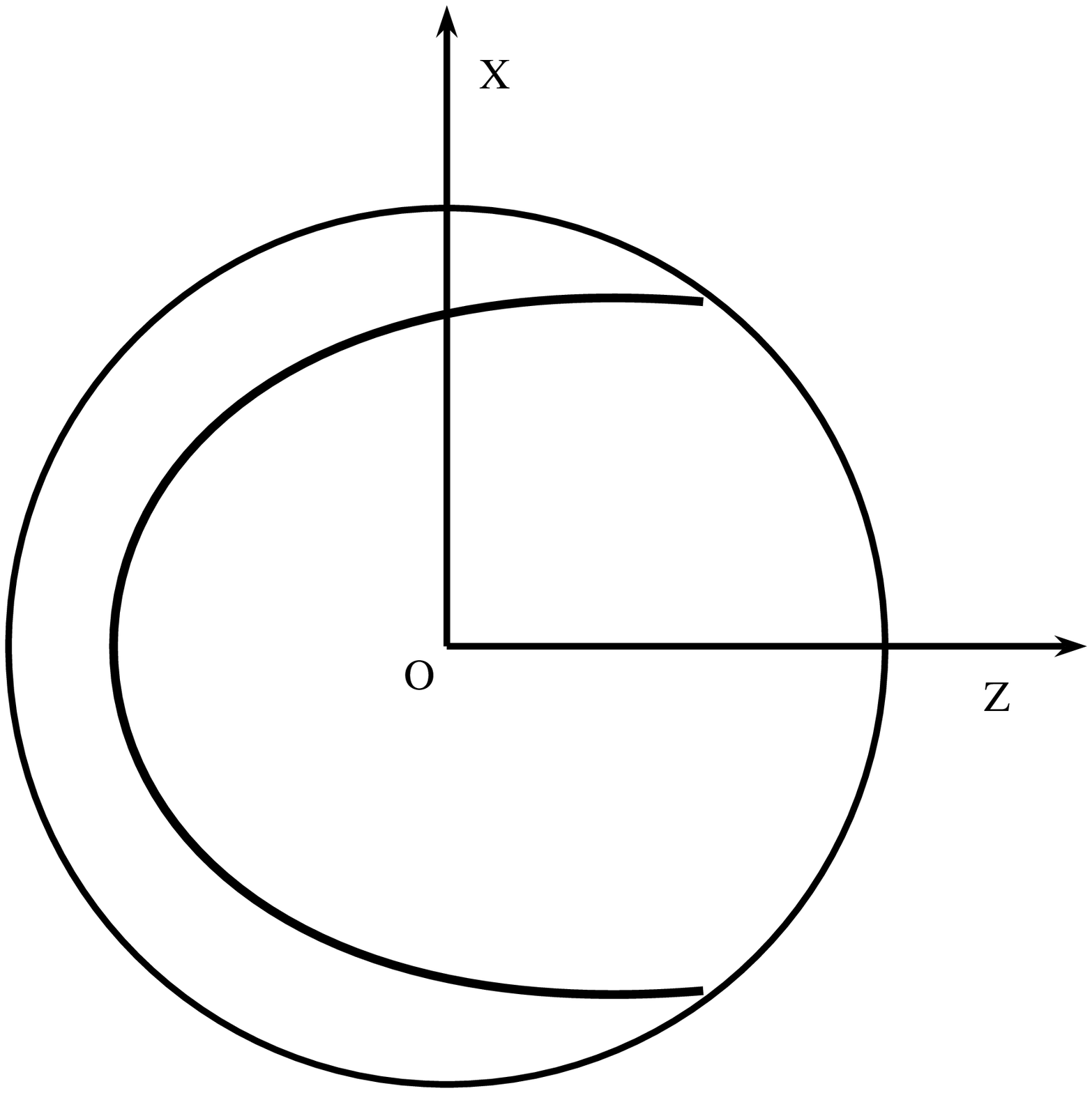}} \hfil \scalebox{0.3}{\includegraphics{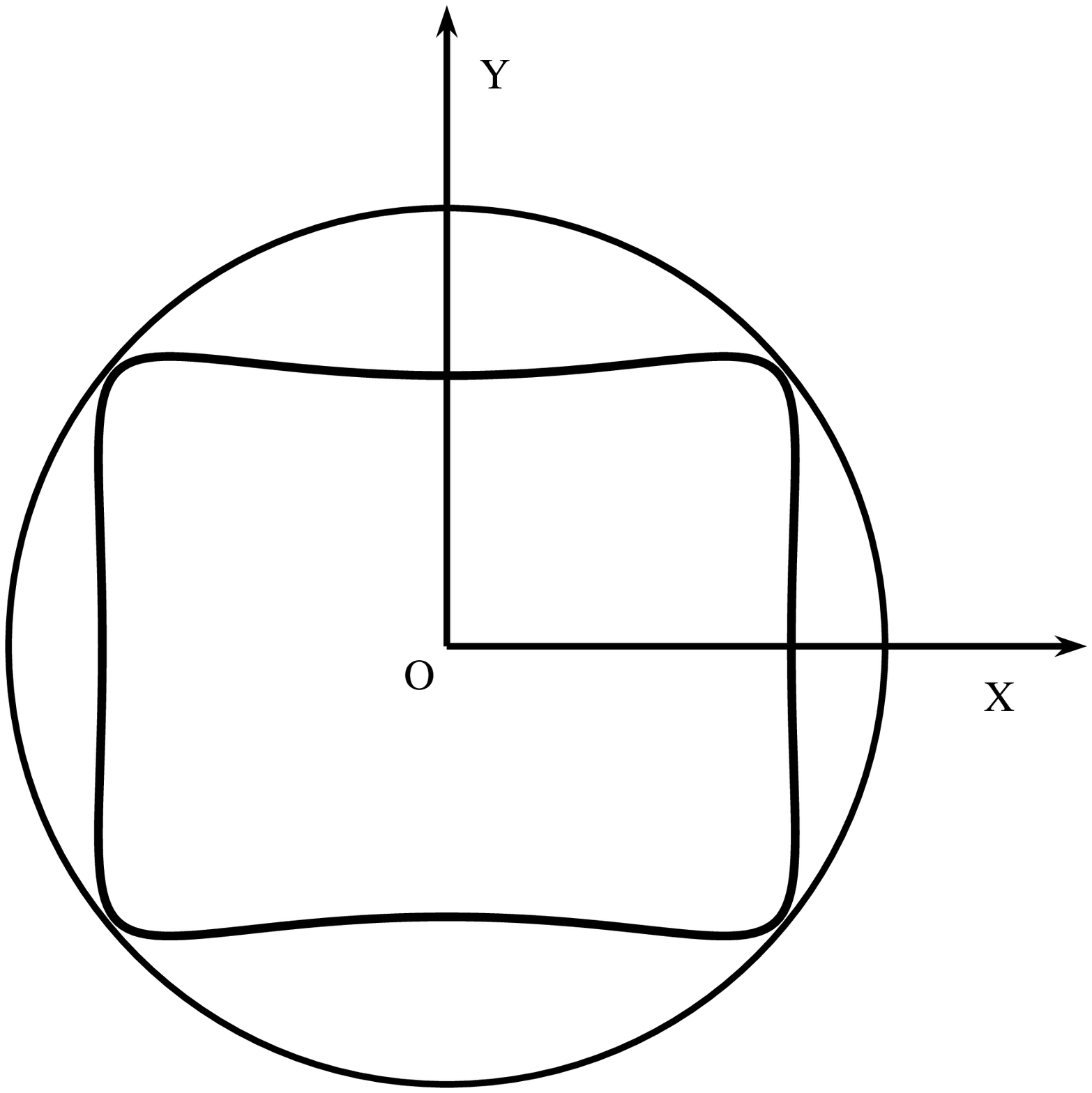}} \hfil

\caption{The trajectory followed by the vector forming the second row of the rotation matrix in a period. The curve is symmetric with respect to the same two
coordinate planes that the body sistem angular momentum vector. Perspective view and projection on the coordinate planes in the system of principal moments of
inertia.}

\end{figure}

One substitutes the equations discovered at page 575, equation 16.21 of the Abramowitz and Stegun handbook \cite{12} for functions $sn$ and $cn$ of complex
argument, which have been written in function of the three components of the unit vector along the angular momentum. These formula were not found in other
classical references and therefore were verified starting from the addition formulae for the Jacobi elliptic functions and the Jacobi equations for the same
functions of imaginary argument which are reproduced in many references.
\begin{equation}
{\bf s} - i {\bf t} = \frac{H(K(k_1)) \Theta(\alpha_1 + i \alpha_2)}{H(K(k_1) + i \alpha_2) \Theta(\alpha_1)} \left( \begin{array}{c} \displaystyle{ \frac{u_2 +
i u_3 u_1}{1 - u_3^2} }\\
\displaystyle{ \frac{-u_1 + i u_2 u_3}{1 - u_3^2} } \\
-i
\end{array} \right)
\end{equation}
We verify in this form the print mistake in the first component of the complex vector which was corrected to be orthogonal both its real part and its imaginary
part to the vector $\bf u$, as it should be.

To establish clearly these equations one rewrites them in terms of Euler angles as defined by Goldstein \cite{13}. Our matrix $\bf R$ corresponds to his
\mbox{\boldmath $\tilde{\rm A}$}. The third row is the unit vector in the angular momentum direction which components are written in terms of two Euler angles
\begin{equation}
{\bf u}^{\rm T} = (u_1 , u_2, u_3) = (\sin \theta \sin \psi, \sin \theta \cos \psi, \cos \theta) \, .
\end{equation}

Adding the first row to the second multiplied by the square root of minus one of the rotation matrix, it is possible to extract a common factor of the three
complex components to find the null vector
\begin{equation}
e^{i \phi} (\cos \psi + i \sin \psi \cos \theta, - \sin \psi + i \cos \psi \cos \theta, - i \sin \theta) \, .
\end{equation}
It is interesting to observe that if the vector between parentheses is divided by $\sin \theta$ it appears again the vector one has found in the Jacobi's
publication
\begin{equation}
\left( \begin{array}{c}
\displaystyle{\frac{\cos \psi}{\sin \theta}} + i \sin \psi \cot \theta \\ [1ex]
 - \displaystyle{\frac{\sin \psi}{\sin \theta}} + i \cos \psi \cot \theta \\
 - i
\end{array} \right) = \left( \begin{array}{c} \displaystyle{ \frac{u_2 + i u_3 u_1}{1 - u_3^2} }\\ [1ex]
\displaystyle{ \frac{-u_1 + i u_2 u_3}{1 - u_3^2} } \\
-i
\end{array} \right) = \left( \begin{array}{c}
{\rm sn}(\alpha_1 + i \alpha_2, k_1) \\ [2ex]
-{\rm cn}(\alpha_1 + i \alpha_2, k_1) \\ [2ex]
-i
\end{array} \right)
\end{equation}

For computations we should use the middle form because it has been written in terms of the components before computed of the unit vector $\bf u$. What means an
economy in computing machine time.

 The factor that should multiply this vector to obtain the two first rows of the rotation matrix as its real and imaginary parts of the vector (42) is $e^{i
 \phi_1} \sin \theta$, where $\phi_1$ is the main periodic part of the third Euler angle $\phi$. According to Jacobi that factor is
$$
e^{i \phi_1} \sqrt{1 - u_3^2}  = \frac{H(K(k_1)) \Theta(\alpha_1 + i \hat{\alpha}_2)}{H(K(k_1) + i \hat{\alpha}_2) \Theta(\alpha_1)} = {\rm
cn}(\hat{\alpha}_2,k_2) \frac{\Theta(0) \Theta(\alpha_1 + i \hat{\alpha}_2)}{\Theta(\alpha_1) \Theta(i \hat{\alpha}_2)} \, .
$$
Actually, ploting the real vs. the imaginary parts of the $\exp i \phi_1$, the points do not fall on a unit circle as it should be. The reason seams subtle to
me. A graphical solution was to use the other $\alpha_2$ root in the right hand side of this equation, which was denoted as $\hat{\alpha}_2$ which is equal to $2
K(k_2) - \alpha_2$. Modification of the right hand side is performed by the prescription 16.33.4 in reference \cite{12} who adds a constant angular velocity
along the angular momentum vector with the same period that vector $\bf u$
$$
e^{i \phi_1} \sqrt{1 - u_3^2}  = \frac{H(K(k_1)) \Theta(\alpha_1 + i \alpha_2)}{H(K(k_1) + i \alpha_2) \Theta(\alpha_1)} \exp\left(\frac{- i \pi \alpha_1}{2
K(k_1)}\right)
$$
\begin{equation}
= {\rm cn}(\alpha_2,k_2) \frac{\Theta(0) \Theta(\alpha_1 + i \alpha_2)}{\Theta(\alpha_1) \Theta(i \alpha_2)} \exp\left(\frac{- i \pi \alpha_1}{2 K(k_1)}\right)\,
.
\end{equation}

This Jacobi's equation allows to obtain the main periodic part of the third Euler angle which remains to know to complete the three Euler angles as a function of
time. In fact one has the real quantities $\sqrt{1-u_3^2}$, ${\rm cn}(\alpha_2,k_2)$, $\Theta(0)$, $\Theta(i \alpha_2)$, $\Theta(\alpha_1)$, therefore taking the
logarithm of (44) and subtracting its complex conjugated expression give us
\begin{equation}
2 i \phi_1 = \ln \frac{\Theta(\alpha_1+i \alpha_2)}{\Theta(\alpha_1 - i \alpha_2)} - i \frac{\pi \alpha_1}{K(k_1)}\, ,
\end{equation}
as is found in references \cite{1} and \cite{14}, as a periodic part with the same period of the $\bf u$ vector. But Jacobi's equation (44) contains more
information since provides with economy of computation, the factor that is needed to compute the two missing rows of the rotation matrix.

The difference of the two angles $\phi$ and $\phi_1$ is an angle with constant angular velocity. This was neglected by Jacobi due to its simplicity. One draws
the figures of the unit vectors forming the Jacobi's rotation matrix neglecting the same constant angular velocity around the constant angular momentum.

In figures 2 and 3 are drawn the curves that follow those two rows of the rotation matrix on the unit sphere in a period $4 K(k_1)$ of $\alpha_1$, for a typical
particular case, computed from equation (44).

\section{Recovering the Jacobi's expression for the components of the rotation matrix}
In this section one proofs the Jacobi's equation (44) of the precedent section which could be used as an efficient algorithm to compute the drawing of the other
curves of two rows of the rotation matrix. The full rotation matrix including the missing constant angular velocity rotation around the constant angular momentum
vector is fully reincorporated here.

One presents a simplified mathematical proof to verify equation (44). As an alternative one could use equation (45) that seems to be more elegant and simple.
However, as has been pointed out, it hides information, it implies first compute the angle, and then compute the trigonometric functions and replace in the
rotation matrix. The Jacobi's equation (44) give directly the trigonometric functions and permits "quick and accurate computation" \cite{6}.

 The derivative of the angle $\phi$ is obtained by Landau and Lifshitz \cite{14} from the angular velocity and angular momentum in terms of the Euler angles (the
 same convention for the Euler angles that used Goldstein \cite{13} and is adopted in this work)
$$
\dot{\phi} = \frac{\omega_1 u_1 + \omega_2 u_2}{1 - u_3^2} = \frac{{\bf u}^{\rm T} \mbox{\boldmath $\omega$} - u_3 \omega_3}{1 - u_3^2} = \frac{2 E/J -u_3^2
J/I_3}{1 - u_3^2} = \frac{J}{I_3} + \frac{J P (e_0 - e_3)}{1 - u_3^2} =
$$
\begin{equation}
\frac{J}{I_3} + J P \sqrt{(e_1-e_0)((e_2-e_3)} \frac{e_0-e_3}{\sqrt{(e_1-e_0)((e_2-e_3)}} \frac{1}{1-u_3^2} \, .
\end{equation}

The first term of the right hand side of this equation is a constant, with an integral lineal in time. The second summand is a periodic function of the variable
$\alpha_1$ with period $4 K(k_1)$, considered by Jacobi. The integral with respect to time of a periodic function is other function lineal in time plus a
periodic function of the same period. The behaviour of $\phi$ is then separated in two terms, one periodic function of the variable $\alpha_1$ of period $4
K(k_1)$ (the main peridic term) plus other function lineal in time, where the two contributions lineal in time are added. The last is a motion with constant
angular velocity, with other period generally incommensurable to the previous one. This last with a constant angular velocity is separated and neglected by
Jacobi due to its relative simplicity. In what follows this terms are not longer ignored.

From  the two terms in the right hand side of (46), the second which will be denoted $d \phi_0/ d t$, is convenient to express in terms of the coordinate
$\alpha_1$
\begin{equation}
\frac{d \phi_0}{d \alpha_1} = \frac{e_0-e_3}{\sqrt{(e_1-e_0)((e_2-e_3)}} \frac{1}{1 - {\rm sn}^2(\alpha_2,k_2) {\rm dn}^2(\alpha_1,k_1)} \, .
\end{equation}

The same derivative is now transformed replacing the constants in terms of elliptic functions of the imaginary value $i \alpha_2$, which is accomplished by means
of our equations (27), and Jacobi's formulas for imaginary argument \cite{7}, \cite{1}.
\begin{equation}
\frac{d \phi_0}{d \alpha_1} = i \frac{{\rm cn}(i \alpha_2,k_1) {\rm dn}(i \alpha_2,k_1)}{{\rm sn}(i \alpha_2,k_1)} \frac{1}{1 - k_1^2 {\rm sn}^2(i \alpha_2,k_1)
{\rm sn}^2(\alpha_1,k_1)} \, .
\end{equation}
In what follows one suppressed the argument $k_1$ since it is the same for all the functions.

It is customary to present the elliptic integral of third kind in the standard form \cite{5}
\begin{equation}
\Pi(\alpha_1, i \alpha_2) = \int_0^{\alpha_1} d \alpha_1 \frac{k_1^2 {\rm sn}(i \alpha_2) {\rm cn}(i \alpha_2) {\rm dn}(i \alpha_2) {\rm sn}^2(\alpha_1)}{1 -
k_1^2 {\rm sn}^2(i \alpha_2) {\rm sn}^2(\alpha_1)} \, ,
\end{equation}
which is close to the angle $\phi_0(\alpha_1)$ because
\begin{equation}
\Pi(\alpha_1,i \alpha_2) = \frac{{\rm cn}(i \alpha_2) {\rm dn}(i \alpha_2)}{{\rm sn}(i \alpha_2)} \left[ \int_0^{\alpha_1} d \alpha_1 \frac{1}{1 - k_1^2 {\rm
sn}^2(i \alpha_2) {\rm sn}^2(\alpha_1)} - \alpha_1 \right] \, .
\end{equation}

One finds so
\begin{equation}
\phi_0(\alpha_1) = i \Pi(\alpha_1, i \alpha_2) + i \frac{{\rm cn}(i \alpha_2) {\rm dn}(i \alpha_2)}{{\rm sn}(i \alpha_2)} \alpha_1 \, ,
\end{equation}
where the constant of integration is assumed to be $\phi_0(0)=0$. However instead of using the elliptic integral of third kind only the Jacobi's Theta functions
are used here.

To obtain the expression of the factor used by Jacobi in his rotation matrix one finds a relation borrowed from the Whittaker and Watson book of analysis
\cite{5}, p 518
\begin{equation}
\frac{\Theta'(\alpha_1 + i \alpha_2)}{\Theta(\alpha_1 + i \alpha_2)} - \frac{\Theta'(\alpha_1)}{\Theta(\alpha_1)} - \frac{\Theta'(i \alpha_2)}{\Theta(i
\alpha_2)} = - k_1^2 {\rm sn}(\alpha_1) {\rm sn}(i \alpha_2) {\rm sn}(\alpha_1 + i \alpha_2) \, .
\end{equation}
Which right hand side is written with the addition formula of Jacobi's function sn$(\alpha_1+i \alpha_2)$ used in (43) as
$$
\frac{\Theta'(\alpha_1 + i \alpha_2)}{\Theta(\alpha_1 + i \alpha_2)} - \frac{\Theta'(\alpha_1)}{\Theta(\alpha_1)} - \frac{\Theta'(i \alpha_2)}{\Theta(i
\alpha_2)} =
$$
$$
\frac{- k_1^2 {\rm sn}(i \alpha_2) {\rm cn}(i \alpha_2) {\rm dn}(i \alpha_2){\rm sn}^2(\alpha_1)}{1 - k_1^2 {\rm sn}^2(\alpha_1) {\rm sn}^2(i \alpha_2)} +
$$
\begin{equation}
\frac{k_1^2 {\rm sn}^2(\alpha_2,k_2) {\rm sn}(\alpha_1,k_1) {\rm cn}(\alpha_1,k_1) {\rm dn}(\alpha_1,k_1)}{1 - {\rm sn}^2(\alpha_2,k_2) {\rm
dn}^2(\alpha_1,k_1)}\, .
\end{equation}

Integrating both sides of this equation from 0 to $\alpha_1$ one finds
$$
\ln \frac{\Theta(0) \Theta(\alpha_1 + i \alpha_2)}{\Theta(\alpha_1) \Theta(i \alpha_2)} -\alpha_1 \frac{\Theta'(i \alpha_2)}{\Theta(i \alpha_2)} =
$$
\begin{equation}
\int_0^{\alpha_1} \frac{- k_1^2 {\rm sn}(i \alpha_2) {\rm cn}(i \alpha_2) {\rm dn}(i \alpha_2){\rm sn}^2(\alpha_1)}{1 - k_1^2 {\rm sn}^2(\alpha_1) {\rm sn}^2(i
\alpha_2)} d\ \alpha_1 + \ln \frac{\sqrt{1-{\rm sn}^2(\alpha_2,k_2) {\rm dn}^2(\alpha_1,k_1)}}{{\rm cn}(\alpha_2,k_2)}\, ,
\end{equation}
where one replaces the angle $\phi_0$ to find a version of the Jacobi equation (44) if one suppresses two of the three terms lineal in the time that are
neglected by Jacobi
\begin{equation}
\ln \frac{\Theta(0) \Theta(\alpha_1 + i \alpha_2)}{\Theta(\alpha_1) \Theta(i \alpha_2)} = i \left[\phi_1(\alpha_1) + \frac{\pi \alpha_1}{2 K(k_1)}\right] + \ln
\frac{\sqrt{1-{\rm sn}^2(\alpha_2,k_2) {\rm dn}^2(\alpha_1,k_1)}}{{\rm cn}(\alpha_2,k_2)}\, ,
\end{equation}
where the lineal terms in time are no longer present as
\begin{equation}
\phi_0(\alpha_1) - \phi_1(\alpha_1) = \frac{\pi \alpha_1}{2 K(k_1)} + i \alpha_1 \left[\frac{\Theta'(i \alpha_2)}{\Theta(i \alpha_2)} + \frac{{\rm cn}(i
\alpha_2) {\rm dn}(i \alpha_2)}{{\rm sn}(i \alpha_2)} \right] \, .
\end{equation}

This constant angular has been ignored in this work, as in Jacobi's, taking in account only the main period of the rotation matrix, in which we have suppressed
such constant angular velocity. One regards preferable this choice as discussed at once.

That constant angular velocity has two terms, one of them depends on parameter $H$ appearing in (20) and disappearing from discussion until reappears as a
summand at the velocity of the angle of Euler $\dot{\phi}$ in (46):
\begin{equation}
\frac{J}{I_3} = J H + J P e_3 \, .
\end{equation}

One adds all the lineal terms in time to have the complete angle $\phi$ minus its periodic $\phi_1$, considered by Jacobi
\begin{equation}
\phi - \phi_1 = J H t + \alpha_1 \left[ \frac{e_0}{\sqrt{(e_1-e_0)(e_2-e_3)}} + \frac{\Theta'(\alpha_2,k_2)}{\Theta(\alpha_2,k_2)} + \frac{\pi
(\alpha_2+K(k_2))}{2 K(k_1) K(k_2)} \right] \, .
\end{equation}

The term proportional to $H$ could have any value and give an undetermined  character to any draw of the rotation matrix, as a consequence it is a vagary to
pretend to draw it. Nevertheless it deserves some extra consideration. The first term and the first inside the parentheses should be reassembled as
\begin{equation}
J H t + \alpha_1 \frac{e_0}{\sqrt{(e_1-e_0)(e_2-e_3)}} = \frac{E}{J} \, t \, .
\end{equation}

In the remaining terms in that Euler angle we see the logarithmic derivative of $\Theta(\alpha_2)$ which is known in analysis as the $Z(\alpha_2)$.
\begin{equation}
\phi - \phi_1 = \frac{E}{J} \, t + \alpha_1 \left[ \frac{\Theta'(\alpha_2,k_2)}{\Theta(\alpha_2,k_2)} + \frac{\pi (\alpha_2+K(k_2))}{2 K(k_1) K(k_2)} \right] \,
.
\end{equation}

 The logarithmic derivative $Z(\alpha_2, k_2) = \Theta'(\alpha_2,k_2)/\Theta(\alpha_2,k_2)$ is computed as the ratio of Fourier series or it is used the
 algorithm of the arithmetic-geometric mean as is found in the literature \cite{12}.

 As one has computed the nine entries of the main period of the rotation matrix, it is possible to draw on the unit sphere other three curves corresponding now
 to the three columns of it. They give the main periodic motion in the inertial system of coordinates of the rotating rigid body with the constant angular
 velocity around the constant angular momentum. The resulting curves are different to the previous result: one finds one curve almost plane symmetric with
 respect to two planes. The two other curves, are both symmetric with respect to one of the two coordinate planes. Transforming one into the other by reflection
 in the other two planes. The three are curves without crossing point.

\section{Drawing the herpolhode}
The equation of energy conservation when the components of the angular momentum and angular velocity vectors  are given in the inertial system, imply that the
component of the angular velocity along the angular momentum is constant
\begin{equation}
2 E/J = {\bf k}^{\rm T} \mbox{\boldmath $\Omega$} \, .
\end{equation}

The components of {\boldmath $\Omega$} perpendicular to $\bf k$ move on the plane perpendicular to $\bf k$, called the invariable plane; its trajectory on the
invariable plane follow the curve called herpolhode. The vector {\boldmath $\Omega$} minus the component of {\boldmath $\Omega$} along the angular momentum $\bf
k$ is written with the double $\times$ product as
\begin{equation}
({\bf k} \times \mbox{\boldmath $\Omega$}) \times {\bf k} \, .
\end{equation}

This vector in the body system is
\begin{equation}
({\bf u} \times \mbox{\boldmath $\omega$}) \times {\bf u} = \dot{\bf u} \times {\bf u} \, ,
\end{equation}
where equation (15) was utilized.

As a consequence the vector describing the herpolhode is
\begin{equation}
{\bf R} (\dot{\bf u} \times {\bf u}) \, .
\end{equation}

When studying the sphero-conal coordinates (24) one finds the tangent vectors to the coordinate lines
\begin{equation}
{\bf e}_1 = \frac{\partial {\bf u}}{\partial \alpha_1} = \left( \begin{array}{c}
- {\rm cn}(\alpha_2, k_2) {\rm sn}(\alpha_1, k_1) {\rm dn}(\alpha_1,k_1) \\
{\rm dn}(\alpha_2, k_2) {\rm cn}(\alpha_1, k_1) {\rm dn}(\alpha_1,k_1) \\
- {\rm sn}(\alpha_2, k_2) k_1^2 {\rm sn}(\alpha_1, k_1) {\rm cn}(\alpha_1,k_1)
\end{array} \right)
\end{equation}
and
\begin{equation}
{\bf e}_2 = \frac{\partial {\bf u}}{\partial \alpha_2} = \left( \begin{array}{c}
- {\rm sn}(\alpha_2, k_2) {\rm dn}(\alpha_2, k_2) {\rm cn}(\alpha_1,k_1) \\
- k_2^2 {\rm sn}(\alpha_2, k_2) {\rm cn}(\alpha_2, k_2) {\rm sn}(\alpha_1,k_1) \\
{\rm cn}(\alpha_2, k_2) {\rm dn}(\alpha_2, k_2) {\rm dn}(\alpha_1,k_1)
\end{array} \right) \, .
\end{equation}
These vectors are perpendicular to vector $\bf u$, and mutually perpendicular. The two vectors are of the same magnitude. Hence one has
\begin{equation}
{\bf e}_1 \times {\bf u} = - {\bf e}_2 \, .
\end{equation}

\begin{figure}

\hfil \scalebox{0.5}{\includegraphics{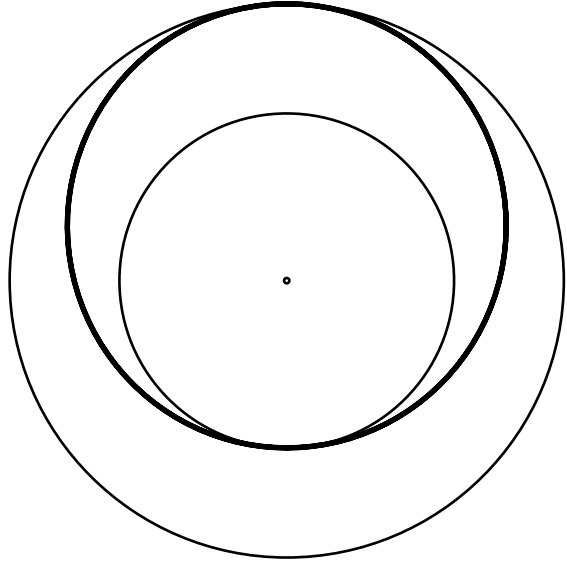}} \hfil

\

\caption{The herpolhode without the term of constant angular velocity of the rotation matrix. It is tangent to two concentric circles.}

\end{figure}

Note vectors $\dot{\bf u}$ and ${\bf e}_1$ are different by the constant factor $\dot{\alpha}_1$, therefore vectors $\dot{\bf u} \times {\bf u}$  and $- {\bf
e}_2$ are different by the same constant factor. Besides, because $\alpha_2$ is a constant coordinate, one makes the substitution of the constants (27) in the
vector ${\bf e}_2$.  one deduces that the entries of the herpolhode $\Omega_1$ and $\Omega_2$ are
\begin{equation}
\left( \begin{array}{c}
\Omega_1 \\
\Omega_2 \\
0
\end{array}
\right) = - \dot{\alpha_1} {\bf R} {\bf e}_2 \, ,
\end{equation}
Eliminating a common constant factor of the components of this vector one deals
\begin{equation}
{\bf R} \left( \begin{array}{c}
\displaystyle{\sqrt{\frac{(e_1 -e_0)(e_2 - e_3)}{e_2 -e_0}}} \, {\rm cn}(\alpha_1,k_1) \\ [2 ex]
\displaystyle{\sqrt{\frac{(e_1-e_3)(e_2-e_0)}{e_1-e_0}}} \, {\rm sn}(\alpha_1,k_1) \\ [2 ex]
- \displaystyle{\sqrt{\frac{(e_0-e_3)(e_2-e_3)}{e_2-e_0}}} \, {\rm dn}(\alpha_1,k_1)
\end{array} \right) \, .
\end{equation}

We can trust to the computer the drawing and computation of the herpolhode until the rotation with constant angular velocity around the angular momentum vector.
In Fig. 4 one finds this curve where one notes the few known properties of the herpolhode. It is symmetric with respect to one of the coordinate axis. We find
the curve between two concentric circles. It is tangent to them at points on the coordinate axis of symmetry. As time increases, the tangent vector to the
herpolhode rotates always in the same direction, until the value $2 \pi$ is reached in a complete period corresponding to $2 K(k_1)$. Some discrepancy is
apparent when comparing with pictures of the herpolhode in the literature \cite{15}, but the motivation to this discrepancy lays in the fact that other authors
do not suppress the action of the constant angular velocity that we ignore. This is an useful addition to the previous work in the literature.


\end{document}